\documentclass[pre,twocolumn,aps,showpacs]{revtex4}
\newcommand{\bec}[1]{\mbox{\boldmath $ #1$}}
\usepackage{graphicx}
\begin{document}
\bigskip
\bigskip
\title{Nonlinear turbulent magnetic diffusion and effective drift
velocity of large-scale magnetic field in a two-dimensional
magnetohydrodynamic turbulence}
\author{Nathan Kleeorin}
\email{nat@menix.bgu.ac.il}
\author{Igor Rogachevskii}
\email{gary@bgu.ac.il} \homepage{http://www.bgu.ac.il/~gary}
\affiliation{Department of Mechanical Engineering, The Ben-Gurion
University of the Negev, POB 653, Beer-Sheva 84105, Israel}
\date{\today}
\begin{abstract}
We study a nonlinear quenching of turbulent magnetic diffusion and
effective drift velocity of large-scale magnetic field in a
developed two-dimensional MHD turbulence at large magnetic Reynolds
numbers. We show that transport of the mean-square magnetic
potential strongly changes quenching of turbulent magnetic
diffusion. In particularly, the catastrophic quenching of turbulent
magnetic diffusion does not occur for the large-scale magnetic
fields $ B \gg B_{\rm eq} / \sqrt{\rm Rm}$ when a divergence of the
flux of the mean-square magnetic potential is not zero, where
$B_{\rm eq}$ is the equipartition mean magnetic field determined by
the turbulent kinetic energy and ${\rm Rm}$ is the magnetic Reynolds
number. In this case the quenching of turbulent magnetic diffusion
is independent of magnetic Reynolds number. The situation is similar
to three-dimensional MHD turbulence at large magnetic Reynolds
numbers whereby the catastrophic quenching of the $\alpha$ effect
does not occur when a divergence of the flux of the small-scale
magnetic helicity is not zero.
\end{abstract}


\maketitle

\section{Introduction}

The magnetic fields of the Sun, solar type stars, galaxies and
planets are believed to be generated by a dynamo process due to the
simultaneous action of the $\alpha$ effect (the helical motions of
turbulence) and differential rotation (see, e.g.,
\cite{M78,P79,KR80,ZRS83,RSS88,S89}). The kinematic stage of the
mean-field dynamo, i.e. the growth of a weak mean magnetic field
with negligible effect on the turbulent flows, is well understood,
while the nonlinear stage of dynamo evolution is a topic of
intensive discussions (for reviews, see \cite{KU99,O03,BS05}). The
most contentious issue is the question of the equilibrium magnetic
field strength at which dynamo action saturates. In particular, the
problem of catastrophic quenching of the $\alpha$ effect in a
developed three-dimensional magnetohydrodynamic (MHD) turbulence
with large magnetic Reynolds numbers has been intensively discussed
in astrophysics and magnetohydrodynamics during last years (see,
e.g., \cite{CV91,VC92,GD94,CH96}). The catastrophic quenching
implies very strong reduction of the $\alpha$ effect during the
growth of the mean magnetic field so that the dynamo generated
magnetic field should be saturated at a very low level. However,
this is in contradiction with observations of the magnetic fields of
the Sun, stars and galaxies.

In a two-dimensional MHD turbulence with imposed large-scale
magnetic field at large magnetic Reynolds numbers, the catastrophic
quenching can occur for turbulent magnetic diffusion (see, e.g.,
\cite{CV91,DHK05}). In particular, small-scale magnetic fluctuations
strongly affect the large-scale magnetic field dynamics even for
very weak mean fields. This causes a strong reduction of turbulent
magnetic diffusion \cite{CV91}. This conclusion is based on
Zeldovich theorem \cite{Z57}. In a two-dimensional MHD turbulence
energy is transferred from large-scale stirring to small scales and
dissipated due to an Alfvenized cascade, whereby eddy energy is
converted to Alfven wave energy (see, e.g., \cite{I64,K65}). The
above discussed quenching is caused by the tendency of the mean
magnetic field to Alfvenize the turbulence.

A principal difference between two-dimensional and three-dimensional
MHD turbulence is related to different integral of motions for these
kind of turbulence. In particular, square of total (small-scale and
large-scale) magnetic potential is conserved in two-dimensional MHD
turbulence, while total (small-scale and large-scale) magnetic
helicity is conserved in three-dimensional MHD turbulence. The
magnetic helicity and the $\alpha$ effect can be positive and
negative, while the squared magnetic potential is only positive. A
comprehensive comparison between two-dimensional and
three-dimensional MHD turbulence has been performed in
\cite{DHK05,DII05}.

It has been recently recognized \cite{BF00,KMRS00} that in
three-dimensional MHD turbulence the catastrophic quenching of the
$\alpha$ effect does not arises when a divergence of the flux of
magnetic helicity is not zero (see also \cite{BS05,KKMR03,RKL06}).
In the present study we show that in a developed two-dimensional MHD
turbulence with large magnetic Reynolds numbers ${\rm Rm}$, a
non-zero divergence of the flux of the mean-square magnetic
potential strongly changes a balance in the equation for these
fluctuations and results in that the catastrophic quenching of
turbulent magnetic diffusion does not occur for the magnetic fields
$B \gg B_{\rm eq} / \sqrt{\rm Rm}$, where $B_{\rm eq}$ is the
equipartition mean magnetic field determined by the turbulent
kinetic energy.

This paper is organized as follows. In Sec.~II we formulate the
governing equations, the assumptions, the procedure of the
derivations. In Sec.~III we determine the nonlinear turbulent
magnetic diffusion coefficients and the nonlinear drift velocities
of the mean magnetic field in a developed two-dimensional MHD
turbulence. Finally, we draw conclusions in Sec.~IV. In Appendix~A
we perform the derivation of the nonlinear turbulent magnetic
diffusion and the nonlinear drift velocities of the mean magnetic
field and in Appendix~B we present the nonlinear functions used in
Sec.~III and their asymptotic formulas.

\section{Governing equations and the procedure of derivation}

Let us consider a developed two-dimensional MHD turbulence with
large hydrodynamic and magnetic Reynolds numbers. We study nonlinear
quenching of the turbulent magnetic diffusion and the effective
drift velocity of the magnetic field. We use a mean field approach
whereby the velocity, pressure and magnetic field are separated into
the mean and fluctuating parts. In a two-dimensional MHD turbulence
the mean magnetic field is ${\bf B} = \bec{\nabla} {\bf \times}
[A(x,y) \, {\bf e}]$, where $A(x,y)$ is the mean magnetic potential
and ${\bf e}$ is the unit vector perpendicular to the plane of the
two-dimensional MHD turbulence, i.e., it is directed along $z$-axis.
The equation for the evolution of the mean magnetic potential for an
incompressible turbulent flow with a zero mean velocity reads:
\begin{eqnarray}
{\partial {A} \over \partial t} + {\rm div} \, \langle {\bf u} \, a
\rangle = \eta \Delta {A} \;, \label{I1}
\end{eqnarray}
where ${\bf u}$ are the velocity fluctuations and $\eta$ is the
magnetic diffusion caused by an electrical conductivity of a fluid.
The mean electromotive force is ${\cal E}_z = \langle {\bf u} \times
{\bf b} \rangle_z = - {\rm div} \, \bec{\Gamma}_a$, where the
spatial flux of magnetic potential $\bec{\Gamma}_a = \langle {\bf u}
\, a \rangle$ and magnetic fluctuations ${\bf b} = \bec{\nabla} {\bf
\times} [a(x,y) \, {\bf e}]$ are described by the fluctuations of
the magnetic potential $a(x,y)$. The mean electromotive force ${\cal
E}_z({\bf B})$ in a two-dimensional MHD turbulence is given by:
\begin{eqnarray}
{\cal E}_{z}({\bf B}) &=& \{[{\bf V}^{\rm eff}({\bf B}) {\bf \times}
{\bf B}]_{i} - {\eta}_{ij}({\bf B}) \, (\bec{\nabla} {\bf \times}
{\bf B})_{j} \} \, e_i  \;, \label{I2}
\end{eqnarray}
where the nonlinear turbulent magnetic diffusion ${\eta}_{ij}({\bf
B})$ and the nonlinear effective drift velocity ${\bf V}^{\rm
eff}({\bf B})$ of the mean magnetic field are determined in Sec.
III.

In order to derive equations for the nonlinear turbulent magnetic
diffusion and the nonlinear effective drift velocity of the mean
magnetic field in a two-dimensional MHD turbulence we use a
procedure outlined below (see Appendix A for details). This
procedure is similar to that used in \cite{RK04} for a study of a
three-dimensional MHD turbulence. We use equations for fluctuations
of velocity and magnetic field
\begin{eqnarray}
{\partial {\bf u}(t,{\bf x}) \over \partial t} &=& - {\bec{\nabla} p
\over \rho} + {1 \over \mu \, \rho} [({\bf b} \cdot \bec{\nabla})
{\bf B} + ({\bf B} \cdot \bec{\nabla}){\bf b}]
\nonumber \\
&& + {\bf u}^N + {\bf F} \,,
\label{B1} \\
{\partial {\bf b}(t,{\bf x}) \over \partial t} &=& ({\bf B} \cdot
\bec{\nabla}){\bf u} - ({\bf u} \cdot \bec{\nabla}) {\bf B} + {\bf
b}^N \,, \label{B2}
\end{eqnarray}
where $\rho$ is the fluid density, ${\bf F}$ is a random external
stirring force, ${\bf u}^{N}$ and ${\bf b}^{N}$ are the nonlinear
terms which include the molecular dissipative terms, $p$ are the
fluctuations of total (hydrodynamic and magnetic) pressure.
Hereafter we omit the magnetic permeability of the fluid $\mu$ and
include $\mu^{-1/2}$ in the definition of magnetic field, we also
omit the density $\rho$ of incompressible fluid and include
$\rho^{1/2}$ in the definition of velocity field. We rewrite
Eqs.~(\ref{B1}) and~(\ref{B2}) in a Fourier space and derive
equations for the two-point second-order correlation functions of
the velocity fluctuations $\langle u_i \, u_j\rangle$, the magnetic
fluctuations $\langle b_i \, b_j \rangle$ and the cross-helicity
$\langle b_i \, u_j \rangle$. The equations for these correlation
functions are given by Eqs.~(\ref{B6})-(\ref{B8}) in Appendix A.

The second-moment equations include the first-order spatial
differential operators $\hat{\cal N}$  applied to the third-order
moments $M^{(III)}$. A problem arises how to close the system, i.e.,
how to express the set of the third-order terms $\hat{\cal N}
M^{(III)}$ through the lower moments $M^{(II)}$ (see, e.g.,
\cite{O70,MY75,Mc90}). We use the spectral $\tau$ approximation
which postulates that the deviations of the third-moment terms,
$\hat{\cal N} M^{(III)}({\bf k})$, from the contributions to these
terms afforded by the background turbulence, $\hat{\cal N}
M^{(III,0)}({\bf k})$, are expressed through the similar deviations
of the second moments, $M^{(II)}({\bf k}) - M^{(II,0)}({\bf k})$:
\begin{eqnarray}
\hat{\cal N} M^{(III)}({\bf k}) &-& \hat{\cal N} M^{(III,0)}({\bf
k})
\nonumber\\
&=& - \, {1 \over \tau(k)} \, [M^{(II)}({\bf k}) - M^{(II,0)}({\bf
k})] \;, \label{AAC3}
\end{eqnarray}
(see, e.g., \cite{O70,PFL76,KRR90,KMR96,RK04}), where $\tau(k)$ is
the scale-dependent relaxation time, which can be identified with
the correlation time of the turbulent velocity field, and the
quantities with the superscript $(0)$ correspond to the background
turbulence. A justification of the $\tau$ approximation for
different situations has been performed in numerical simulations and
analytical studies in \cite{BS05,BF02,FB02,BK04,BSM05,SSB07}.

Next, we split all second-order correlation functions, $M^{(II)}$,
into symmetric and antisymmetric parts with respect to the wave
vector ${\bf k}$. We assume that the characteristic time of
variation of the mean magnetic field ${\bf B}$ is substantially
larger than the correlation time $\tau(k)$ for all turbulence
scales. This allows us to get a stationary solution for the
equations for the second-order moments, $M^{(II)}$. We use a model
of the background anisotropic and inhomogeneous two-dimensional MHD
turbulence  determined by Eqs.~(\ref{K1})-(\ref{K2}) in Appendix A.

In this study we consider an intermediate nonlinearity which implies
that the mean magnetic field is not enough strong in order to affect
the correlation time of turbulent velocity field. The theory for a
very strong mean magnetic field can be modified after taking into
account a dependence of the correlation time of the turbulent
velocity field on the mean magnetic field.

Using the solution of the derived second-moment equations, we
determine the mean electromotive force, ${\cal E}_{i} =
\varepsilon_{imn} \, \int \langle b_n \, u_m \rangle_{\bf k} \,{\rm
d} {\bf k}$ (see Appendix A for details), where $\varepsilon_{ijk}$
is the fully antisymmetric Levi-Civita tensor. This procedure allows
us to determine the nonlinear turbulent magnetic diffusion and the
nonlinear effective drift velocity of the mean magnetic field in a
two-dimensional MHD turbulence.

\section{Turbulent transport coefficients}

The derivation outlined in Sec. II yields the nonlinear turbulent
magnetic diffusion of the mean magnetic field. In particular, in
order to determine the nonlinear turbulent magnetic diffusion
${\eta}_{ij}({\bf B})$ we use an identity: $\eta_{ij} =
(\varepsilon_{ikp} b_{jkp} + \varepsilon_{jkp} b_{ikp}) / 4$, where
the tensor $b_{ijk}$ is determined by Eq.~(\ref{C2}) in Appendix A.
The nonlinear turbulent magnetic diffusion coefficient along the
mean magnetic field, $\eta_{_{B}}$, and the cross-field turbulent
magnetic diffusion coefficient, $\eta_{_{\perp}}$, are given by:
\begin{eqnarray}
\eta_{_{B}} &=& \tau_0 \, [\langle {\bf u}^2 \rangle^{(0)} - \langle
{\bf b}^2 \rangle^{(0)}] \, \Psi_1(\beta)  \;,
\label{BN36}\\
\eta_{_{\perp}} &=& \tau_0 \, [ \langle {\bf u}^2 \rangle^{(0)} -
\langle {\bf b}^2 \rangle^{(0)}] \, \Psi(\beta) \;,
\label{BN37}
\end{eqnarray}
where $\tau_{0} = l_{0} / u_{0}$ and $u_{0}= \sqrt{\langle {\bf u}^2
\rangle^{(0)}} $ is the characteristic turbulent velocity in the
maximum scale of turbulent motions $l_{0}$. The quantities with the
superscript $(0)$ correspond to the background turbulence. The
functions $\Psi(\beta)$, $\, \Psi_1(\beta)$ and their asymptotic
formulas are given in Appendix B, $\beta = 4 \, ( B / B_{\rm eq})$
and $B_{\rm eq} = \sqrt{\langle {\bf u}^2 \rangle^{(0)}}$ is the
equipartition field. More general equations for $\eta_{_{B}}$ and
$\eta_{_{\perp}}$  in the case of an anisotropic background
turbulence are given by Eqs.~(\ref{B36}) and~(\ref{B37}) in Appendix
A. It follows from Eqs.~(\ref{BN36}) and~(\ref{BN37}) that in the
case of Alfvenic equipartition, $\langle {\bf u}^2 \rangle^{(0)} =
\langle {\bf b}^2 \rangle^{(0)}$, the nonlinear turbulent magnetic
diffusion vanishes.

The nonlinear turbulent magnetic diffusion depends on a flux of
mean-square magnetic potential. This flux can change properties of
the quenching of the cross-field turbulent magnetic diffusion.
Indeed, let us determine the parameter $\epsilon = \langle {\bf b}^2
\rangle^{(0)} / \langle {\bf u}^2 \rangle^{(0)}$ using budget
equation for the evolution of the mean-square magnetic potential
$\langle a^2 \rangle$:
\begin{eqnarray}
{\partial \langle a^2 \rangle \over \partial t} + {\rm div} \, {\bf
F}^A = 2 \eta_{_{\perp}} {\bf B}^2 - 2 \eta \langle {\bf b}^2
\rangle \;,
\label{DB32}
\end{eqnarray}
where the flux ${\bf F}^A = \langle {\bf u} \, a^2 \rangle - \eta
\bec{\nabla} \langle a^2 \rangle$ determines the transport of
$\langle a^2 \rangle$. The first term $2 \eta_{_{\perp}} {\bf B}^2$
in the right hand side of Eq.~(\ref{DB32}) describes a production of
the mean-square magnetic potential $\langle a^2 \rangle$, while the
term $- 2 \eta \langle {\bf b}^2 \rangle$ determines the resistive
dissipation of $\langle a^2 \rangle$. In the absence of the flux of
the mean-square magnetic potential, ${\bf F}^A = 0$,
Eq.~(\ref{DB32}) implies the catastrophic quenching of the
cross-field turbulent magnetic diffusion. In particular, in a
steady-state Eq.~(\ref{DB32}) reads $\eta_{_{\perp}} = \eta \,
\langle {\bf b}^2 \rangle / {\bf B}^2$. Since the magnetic energy is
less than the kinetic energy, $\langle {\bf b}^2 \rangle < \langle
{\bf u}^2 \rangle^{(0)}$, we get
\begin{eqnarray}
{\eta_{_{\perp}} \over \eta_{_{T}}} < {1 \over {\rm Rm} \, (B
/B_{\rm eq})^2} \;, \label{DB33}
\end{eqnarray}
where $\eta_{_{T}} = l_0 \, u_{0} / 2$ and ${\rm Rm} = u_0 \, l_0 /
\eta$ is the magnetic Reynolds number. This estimate implies a
strong quenching of the cross-field turbulent magnetic diffusion
with increasing ${\rm Rm} \, (B /B_{\rm eq})^2$ due to Alfvenization
of turbulence by tangling of a weak mean magnetic field by velocity
fluctuations \cite{DHK05}.

Situation is drastically changed when ${\rm div} \, {\bf F}^A \not
=0$. Indeed, Eq.~(\ref{DB32}) is not closed because it depends on
the magnetic energy $\langle {\bf b}^2 \rangle$. The energy of
magnetic fluctuations $\langle {\bf b}^2 \rangle$ can be determined
in the same way as we derived the cross-helicity tensor. In
particular, $\langle {\bf b}^2 \rangle$ is obtained from
Eq.~(\ref{B24}) given in Appendix A, after the integration in $ {\bf
k} $-space. The result is given by
\begin{eqnarray}
\langle {\bf b}^2 \rangle &=& {1 \over 2} \Big[\langle {\bf u}^2
\rangle^{(0)} [1 - \phi(\beta)] + \langle {\bf b}^2 \rangle^{(0)} [1
+ \phi(\beta)] \Big] \;,
\nonumber \\
\label{DDBN33}
\end{eqnarray}
where the function $\phi(\beta)$ and its asymptotic formulas are
given in Appendix B. More general equation for $\langle {\bf b}^2
\rangle$ for anisotropic background turbulence is given by
Eq.~(\ref{DDB33}) in Appendix A.

Equation~(\ref{DB32}) allows us to determine the energy of magnetic
fluctuations of the background turbulence self-consistently. In
particular, combining Eq.~(\ref{DDBN33}) with the steady-state
solution of Eq.~(\ref{DB32}) we determine the parameter $\epsilon =
\langle {\bf b}^2 \rangle^{(0)} / \langle {\bf u}^2 \rangle^{(0)}$
[see, e.g., Eq.~(\ref{R41}) in Appendix A for anisotropic background
turbulence]. When $B \gg B_{\rm eq} / \sqrt{\rm Rm}$, the parameter
$\epsilon$ is given by
\begin{eqnarray}
\epsilon = 1 - {{\rm div} \, {\bf F}^A \over 4 \eta_{_{T}} B^2
\Psi(\beta)} \; .
\label{R42}
\end{eqnarray}
Therefore, Eqs.~(\ref{BN36}), (\ref{BN37}) and~(\ref{R42}) yield the
nonlinear turbulent magnetic diffusion in two directions:
\begin{eqnarray}
\eta_{_{B}} &=& {{\rm div} \, {\bf F}^A \over 2  B^2} \,
\biggl({\Psi_1(\beta) \over \Psi(\beta)} \biggr) \;,
\label{DBN34}\\
\eta_{_{\perp}} &=& {{\rm div} \, {\bf F}^A \over 2  B^2} \; .
\label{DBN35}
\end{eqnarray}
where $\eta_{_{B}}$ is the nonlinear turbulent magnetic diffusion
along the mean magnetic field and $\eta_{_{\perp}}$ is the
cross-field nonlinear turbulent magnetic diffusion. Remarkably,
Eq.~(\ref{DBN35}) can be obtained directly from Eq.~(\ref{DB32})
written in a steady-state if we neglect the resistive dissipation
term $- 2 \eta \langle {\bf b}^2 \rangle$ in the right hand side of
Eq.~(\ref{DB32}).

In order to determine the parameter $\epsilon$ we use the
steady-state solution of Eq.~(\ref{DB32}). However, the steady-state
solution of this equation exists not for all values of the mean
magnetic field. Indeed, let us plot in Fig.~1 the function
$\Psi(\beta)$ for the exponent of the energy spectrum of the
background turbulence $q=5/3$. At $B \to 0.18 \, B_{\rm eq}$ the
function $\Psi(\beta)$ tends to zero (see Fig.~1). In the range $B
\geq 0.18 \, B_{\rm eq}$ the steady-state solution of
Eq.~(\ref{DB32}) does not exist. The turbulent magnetic diffusion
should be positive, which implies that ${\rm div} \, {\bf F}^A > 0$.
Therefore, when ${\rm div} \, {\bf F}^A < 0$ there is no
steady-state solution of Eq.~(\ref{DB32}) for $B \geq 0.18 \, B_{\rm
eq}$ as well. More detailed discussion of this facet is given in
Appendix A after Eq.~(\ref{DB35}).

\begin{figure}
\centering
\includegraphics[width=8cm]{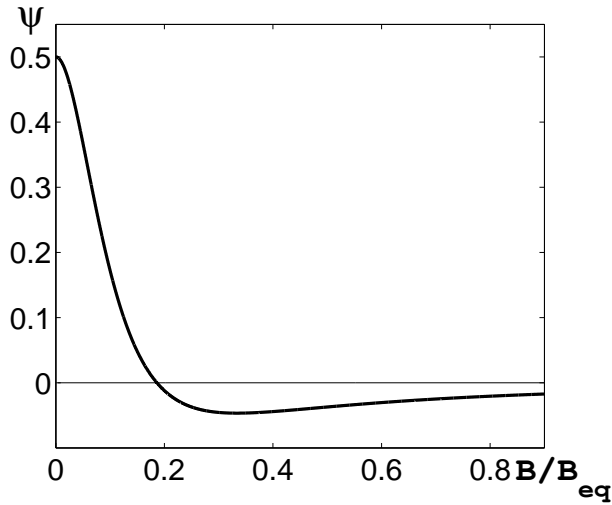}
\caption{\label{Fig1} The function $\Psi(B/B_{\rm eq})$ for
$q=5/3$.}
\end{figure}

In inhomogeneous turbulence there are also turbulent diamagnetic and
paramagnetic effects. In particular, an inhomogeneity of the
velocity fluctuations leads to a transport of mean magnetic flux
from regions with high intensity of the velocity fluctuations
(turbulent diamagnetism, see, e.g., \cite{Z57,KR80}). On the other
hand, an inhomogeneity of magnetic fluctuations due to the
small-scale dynamo causes turbulent paramagnetic velocity, i.e., the
magnetic flux is pushed into regions with high intensity of the
magnetic fluctuations (see, e.g., \cite{VK83,RKR03}). In order to
determine the nonlinear turbulent diamagnetic and paramagnetic drift
velocities ${\bf V}^{\rm eff}({\bf B})$ of the mean magnetic field,
we use an identity: $V^{\rm (eff)}_{k} = \varepsilon_{kji} a_{ij} /
2$, where the tensor $a_{ij}$ is determined by Eq.~(\ref{C1}) in
Appendix A. The inhomogeneities of the velocity and magnetic
fluctuations of the background turbulence are characterized by the
following parameters $\Lambda_i^{(v)} = \nabla_i \langle {\bf u}^2
\rangle^{(0)} / \langle {\bf u}^2 \rangle^{(0)}$ and
$\Lambda_i^{(b)} = \nabla_i \langle {\bf b}^2 \rangle^{(0)} /
\langle {\bf b}^2 \rangle^{(0)}$. The nonlinear effective drift
velocity of the mean magnetic field is given by
\begin{eqnarray}
{\bf V}^{\rm eff} = - 2 \eta_{_{T}} [{\bf \Lambda}^{(v)} - \epsilon
{\bf \Lambda}^{(m)}] \Psi_1(\beta) \;,
\label{DBN40}
\end{eqnarray}
where the function $\Psi_1(\beta)$ and its asymptotic formulas are
given in Appendix B. When $B \gg  B_{\rm eq} / \sqrt{\rm Rm}$,
Eqs.~(\ref{R42}) and~(\ref{DBN40}) yield
\begin{eqnarray}
{\bf V}^{\rm eff} &=& - 2 \eta_{_{T}} \biggl[{\bf \Lambda}^{(v)} -
{\bf \Lambda}^{(m)} + {{\rm div} \, {\bf F}^A \over 4 \eta_{_{T}}
B^2 \Psi(\beta)} \, {\bf \Lambda}^{(m)} \biggr] \Psi_1(\beta) \; .
\nonumber \\
\label{RN43}
\end{eqnarray}
The first term $\propto {\bf \Lambda}^{(v)}$ in Eq.~(\ref{RN43})
determines the turbulent diamagnetic drift velocity while the second
term $\propto {\bf \Lambda}^{(m)}$ describes the turbulent
paramagnetic drift velocity. The last term $\propto {\rm div} \,
{\bf F}^A$ in Eq.~(\ref{RN43}) determines the turbulent diamagnetic
drift velocity caused by magnetic fluctuations for $B < 0.18 \,
B_{\rm eq}$. More general equation for ${\bf V}^{\rm eff}$ for
anisotropic background turbulence is given by Eq.~(\ref{R43}) in
Appendix A.

\section{Conclusions}

In the present study we investigate nonlinear quenching of the
turbulent magnetic diffusion and the effective drift velocity of the
magnetic field in a developed two-dimensional MHD turbulence at
large magnetic Reynolds numbers. We elucidate an important role of
transport of the mean-square magnetic potential which strongly
changes quenching properties of turbulent magnetic diffusion. In
particular, we show that the catastrophic quenching of turbulent
magnetic diffusion does not arises for the magnetic fields $ B \gg
B_{\rm eq} / \sqrt{\rm Rm}$ for a non-zero divergence of the flux of
the mean-square magnetic potential. In this case the quenching of
turbulent magnetic diffusion is independent of magnetic Reynolds
number. This is similar to a three-dimensional MHD turbulence at
large magnetic Reynolds numbers whereby the catastrophic quenching
of the $\alpha$ effect does not occur when a divergence of the flux
of the small-scale magnetic helicity is not zero. Note that in a
two-dimensional MHD turbulence, the magnetic field may only decay,
while in three-dimensional MHD turbulence magnetic field may grow by
dynamo mechanism.

Note that a quenching of turbulent magnetic diffusivity in a 'wavy'
magnetohydrodynamic turbulence in two dimensions was recently
studied in \cite{SD07}. They found that the turbulent  magnetic
diffusivity in the fourth-order does not vanish when the magnetic
Reynolds number tends to infinity. In particularly, the second-order
(quasi-linear) contribution to the spatial flux of the mean magnetic
potential is quenched as ${\rm Rm}^{-1}$, while the fourth-order
contribution to the flux is independent of ${\rm Rm}$. This implies
that the turbulent magnetic diffusivity is not quenched
catastrophically in the presence of dispersive waves which can
transfer the mean-square magnetic potential. These findings are in
an agrement with our results.

\begin{acknowledgments}
We have benefited from stimulating discussions with P. H. Diamond,
who initiated this work during our visit to the Isaac Newton
Institute for Mathematical Sciences (Cambridge) in the framework of
the programme "Magnetohydrodynamics of Stellar Interiors".
\end{acknowledgments}

\appendix

\section{Derivations of the nonlinear turbulent transport coefficients}

We use equations for fluctuations of velocity and magnetic field
written in a Fourier space and derive equations for the second
moments in two-dimensional MHD turbulence using a procedure which is
similar to that used in \cite{RK04} for a study of a
three-dimensional MHD turbulence. In order to exclude the pressure
term from the equation of motion~(\ref{B1}) we determine $
\bec{\nabla} {\bf \times} (\bec{\nabla} {\bf \times} {\bf u}) .$ We
also apply the two-scale approach, e.g., we use large-scale $ {\bf
R} $, $ {\bf K} $ and small-scale ${\bf r}$, ${\bf k}$ variables
(see, e.g., \cite{RS75}). We assume that there exists a separation
of scales, i.e., the maximum scale of turbulent motions $l_0$ is
much smaller then the characteristic scale $L_B$ of inhomogeneities
of the mean magnetic field. We derive equations for the following
correlation functions: $f_{ij}({\bf k, R}) = \hat L(u_i; u_j)$, $\;
h_{ij}({\bf k, R}) = \hat L(b_i; b_j)$ and $g_{ij}({\bf k, R}) =
\hat L(b_i; u_j)$, where
\begin{eqnarray*}
\hat L(a; c) = \int \langle a({\bf k} + {\bf  K} / 2) c(-{\bf k} +
{\bf  K} / 2) \rangle
\\
\times \exp{(i {\bf K} {\bf \cdot} {\bf R}) } \,d {\bf  K} \; .
\end{eqnarray*}
The equations for these correlation functions are given by
\begin{eqnarray}
{\partial f_{ij}({\bf k}) \over \partial t} &=& i({\bf k} {\bf
\cdot} {\bf B}) \Phi_{ij}^{(M)} + I^f_{ij} + F_{ij} + \hat{\cal N}
f_{ij} \;,
\label{B6} \\
{\partial h_{ij}({\bf k}) \over \partial t} &=& - i({\bf k}{\bf
\cdot} {\bf B}) \Phi_{ij}^{(M)} + I^h_{ij} + \hat{\cal N} h_{ij} \;,
\label{B7} \\
{\partial g_{ij}({\bf k }) \over \partial t} &=& i({\bf k} {\bf
\cdot} {\bf B}) [f_{ij}({\bf k}) - h_{ij}({\bf k})] + I^g_{ij} +
\hat{\cal N} g_{ij} \;,
\nonumber\\
\label{B8}
\end{eqnarray}
where hereafter we omit arguments $t$ and ${\bf R}$ in the
correlation functions and neglect terms $ \sim O(\nabla^2) .$ Here
$\Phi_{ij}^{(M)}({\bf k}) = g_{ij}({\bf k}) - g_{ji}(-{\bf k}) ,$ $
\, F_{ij}({\bf k}) = \langle \tilde F_i ({\bf k}) u_j(-{\bf k})
\rangle + \langle u_i({\bf k}) \tilde F_j(-{\bf k })\rangle ,$ and
${\bf \tilde F} ({\bf k}) = {\bf k} {\bf \times} ({\bf k} {\bf
\times} {\bf F}({\bf k})) / k^2 \rho .$ The stirring force ${\bf
F}({\bf k})$ is an external parameter, that determines the
background turbulence. The source terms $I_{ij}^f$ , $\, I_{ij}^h$
and $I_{ij}^g$ which contain the large-scale spatial derivatives of
the mean magnetic field and turbulence are given by
\begin{eqnarray}
I_{ij}^f &=& {1 \over 2}({\bf B} {\bf \cdot} \bec{\nabla})
\Phi_{ij}^{(P)} + [g_{qj}({\bf k}) (2 P_{in}^{(2)}(k) -
\delta_{in}^{(2)})
\nonumber \\
&& + g_{qi}(-{\bf k}) (2 P_{jn}^{(2)}(k) - \delta_{jn}^{(2)})]
B_{n,q} - B_{n,q} k_{n} \Phi_{ijq}^{(P)}\;,
\nonumber\\
\label{M1}\\
I_{ij}^h &=& {1 \over 2}({\bf B} {\bf \cdot} \bec{\nabla})
\Phi_{ij}^{(P)} - [g_{iq}({\bf k}) \delta_{jn}^{(2)} + g_{jq}(-{\bf
k}) \delta_{in}^{(2)}] B_{n,q}
\nonumber\\
&& - B_{n,q} k_{n} \Phi_{ijq}^{(P)} \;,
\label{M2}\\
I_{ij}^g &=& {1 \over 2} ({\bf B} {\bf \cdot} \bec{\nabla}) (f_{ij}
+ h_{ij}) + h_{iq} (2 P_{jn}^{(2)}(k) - \delta_{jn}^{(2)}) B_{n,q}
\nonumber \\
&& - f_{nj} B_{i,n} - B_{n,q} k_{n}(f_{ijq} + h_{ijq}) \;,
\label{M3}
\end{eqnarray}
where $P_{ij}^{(2)}(k) = \delta_{ij}^{(2)} - k_{ij}$, $\, k_{ij} =
k_i k_j / k^2 $, $\, \Phi_{ij}^{(P)}({\bf k}) = g_{ij}({\bf k}) +
g_{ji}(-{\bf k}) ,$ and $B_{i,j} = \nabla_j B_{i} ,$ the terms $
\hat{\cal N} f_{ij}$, $\, \hat{\cal N}h_{ij}$ and $\, \hat{\cal
N}g_{ij}$ are the third-order moment terms appearing due to the
nonlinear terms,  $f_{ijq} = (1/2) \partial f_{ij} / \partial k_{q}
$, and similarly for $h_{ijq}$ and $\Phi_{ijq}^{(P)}$. For the
derivation of Eqs.~(\ref{B6})-(\ref{B8}) we use identities given in
\cite{RK04}. We take into account that in Eq. (\ref{B8}) the terms
with symmetric tensors with respect to the indexes "i" and "j" do
not contribute to the mean electromotive force because ${\cal E}_{m}
= \varepsilon_{mji} \, g_{ij} $.

We use the spectral $\tau$ approximation which postulates that the
deviations of the third-moment terms, $\hat{\cal N} M^{(III)}({\bf
k})$, from the contributions to these terms afforded by the
background turbulence, $\hat{\cal N} M^{(III,0)}({\bf k})$, are
expressed through the similar deviations of the second moments,
$M^{(II)}({\bf k}) - M^{(II,0)}({\bf k})$ [see Eq.~(\ref{AAC3})].
The superscript $ {(0)} $ corresponds to the background turbulence.
First, we solve Eqs.~(\ref{B6})-(\ref{B8}) neglecting the sources
$I^f_{ij}, I^h_{ij}, I^g_{ij}$ with the large-scale spatial
derivatives. Then we will take into account the terms with the
large-scale spatial derivatives by perturbations. We assume that
$\eta k^2 \ll \tau^{-1}(k)$ and $\nu k^2 \ll \tau^{-1}(k)$ for the
inertial range of turbulent flow, where $\nu$ is the kinematic
viscosity and $\eta$ is the magnetic diffusion due to the electrical
conductivity of fluid. We also assume that the characteristic time
of variation of the mean magnetic field ${\bf B}$ is substantially
larger than the correlation time $\tau(k)$ for all turbulence
scales. We split all correlation functions into symmetric and
antisymmetric parts with respect to the wave number ${\bf k}$, {\em
e.g.,} $ f_{ij} = f_{ij}^{(s)} + f_{ij}^{(a)} ,$ where $
f_{ij}^{(s)} = [f_{ij}({\bf k}) + f_{ij}(-{\bf k})] / 2 $ is the
symmetric part and $ f_{ij}^{(a)} = [f_{ij}({\bf k}) - f_{ij}(-{\bf
k})] / 2 $ is the antisymmetric part, and similarly for other
tensors. Thus, in a steady-state Eqs. (\ref{B6})-(\ref{B8}) yield:
\begin{eqnarray}
\hat f_{ij}^{(s)}({\bf k}) &\approx& {1 \over 1 + 2 \psi} [(1 +
\psi) f_{ij}^{(0)}({\bf k}) + \psi h_{ij}^{(0)}({\bf k})]  \;,
\nonumber\\
\label{B22} \\
\hat h_{ij}^{(s)}({\bf k}) &\approx& {1 \over 1 + 2 \psi} [\psi
f_{ij}^{(0)}({\bf k}) + (1 + \psi) h_{ij}^{(0)}({\bf k})] \;,
\nonumber\\
\label{B24}\\
\hat g_{ij}^{(a)}({\bf k}) &\approx& {i \tau ({\bf k} {\bf \cdot}
{\bf B}) \over 1 + 2 \psi} [f_{ij}^{(0)}({\bf k}) -
h_{ij}^{(0)}({\bf k})]  \;,
\label{B26}
\end{eqnarray}
where $ \psi({\bf k}) = 2 (\tau \, {\bf k} {\bf \cdot} {\bf B})^2$,
$\, \hat f_{ij}, \hat h_{ij}$ and $\hat g_{ij}$ are solutions
without the sources $I^f_{ij}, I^h_{ij}$ and $\, I^g_{ij}$.  The
correlation functions $\hat f_{ij}^{(a)}({\bf k})$, $\hat
h_{ij}^{(a)}({\bf k})$ and $\hat g_{ij}^{(s)}({\bf k})$ vanish if we
neglect the large-scale spatial derivatives, i.e., they are
proportional to the first-order spatial derivatives.

Next, we take into account the large-scale spatial derivatives in
Eqs. (\ref{B6})-(\ref{B8}) by perturbations. Their effect determines
the following steady-state equations for the second moments $\tilde
f_{ij}$, $\tilde h_{ij}$ and $\tilde g_{ij}$:
\begin{eqnarray}
\tilde f_{ij}^{(a)}({\bf k}) &=& i \tau ({\bf k} {\bf \cdot} {\bf
B}) \tilde \Phi_{ij}^{(M,s)}({\bf k}) + \tau I^f_{ij} \;,
\label{B28}\\
\tilde h_{ij}^{(a)}({\bf k}) &=& - i \tau ({\bf k} {\bf \cdot} {\bf
B}) \tilde \Phi_{ij}^{(M,s)}({\bf k}) + \tau I^h_{ij} \;,
\label{B29} \\
\tilde g_{ij}^{(s)}({\bf k }) &=& i \tau ({\bf k} {\bf \cdot} {\bf
B}) (\tilde f_{ij}^{(a)}({\bf k}) - \tilde h_{ij}^{(a)}({\bf k})) +
\tau I^g_{ij} \;,
\nonumber \\
\label{B30}
\end{eqnarray}
where $ \tilde \Phi_{ij}^{(M,s)} = [\tilde \Phi_{ij}^{(M)}({\bf k})
+ \tilde \Phi_{ij}^{(M)}(-{\bf k})] / 2 .$ The solution of Eqs.
(\ref{B28})-(\ref{B30}) yield
\begin{eqnarray}
\tilde \Phi_{ij}^{(M,s)}({\bf k}) &=& {\tau \over 1 + 2 \psi} \{
I^g_{ij} - I^g_{ji} + i \tau ({\bf k} {\bf \cdot} {\bf B}) (I^f_{ij}
- I^f_{ji}
\nonumber\\
& & + I^h_{ji} - I^h_{ij}) \} \; . \label{B31}
\end{eqnarray}
Substituting Eq.~(\ref{B31}) into Eqs. (\ref{B28})-(\ref{B30}) we
obtain the final expressions in ${\bf k}$-space for the tensors
$\tilde f_{ij}^{(a)}({\bf k})$, $\tilde h_{ij}^{(a)}({\bf k})$,
$\tilde g_{ij}^{(s)}({\bf k})$ and $\tilde \Phi_{ij}^{(M,s)}({\bf
k})$. In particular,
\begin{eqnarray}
&& \tilde \Phi_{mn}^{(M,s)}({\bf k}) = {\tau(k) W(k) \langle {\bf
u}^2 \rangle^{(0)} \over (1 + 2 \psi)^2} \biggl[(1 + \epsilon)(1 + 2
\psi) (\delta_{nj}^{(2)} \delta_{mk}^{(2)}
\nonumber\\
& & - \delta_{mj}^{(2)} \delta_{nk}^{(2)} + k_{nk} \delta_{mj}^{(2)}
- k_{mk} \delta_{nj}^{(2)}) - 2 (\epsilon + 2 \psi) (k_{nj}
\delta_{mk}^{(2)}
\nonumber\\
& & - k_{mj} \delta_{nk}^{(2)}) \biggr] \,  B_{j,k} \; .
\label{BB40}
\end{eqnarray}
The correlation functions $\tilde f_{ij}^{(s)}({\bf k})$, $\tilde
h_{ij}^{(s)}({\bf k})$ and $\tilde g_{ij}^{(a)}({\bf k})$ are of the
order of $\sim O(\nabla^2)$, i.e., they are proportional to the
second-order spatial derivatives. Thus $\hat g_{ij} + \tilde g_{ij}
$ is the correlation function of the cross-helicity, and similarly
for other second moments. Now we calculate the mean electromotive
force $ {\cal E}_{i}({\bf r}=0) = (1/2) \varepsilon_{inm} \int
\tilde \Phi_{mn}^{(M,s)}({\bf k}) \,d {\bf k} $. Thus,
\begin{eqnarray}
{\cal E}_{i} &=& \varepsilon_{inm} \int {\tau \over 1 + 2 \psi} \,
[I^g_{mn} + i \tau ({\bf k} {\bf \cdot} {\bf B}) (I^f_{mn} -
I^h_{mn})] \,d {\bf k} \; .
\nonumber \\
\label{B35}
\end{eqnarray}

We use the following model of the background anisotropic and
inhomogeneous two-dimensional MHD turbulence:
\begin{eqnarray}
&& \langle u_i \, u_j \rangle^{(0)}({\bf k}) = \langle {\bf u}^2
\rangle^{(0)} \, W^v(k) \, \biggl\{ (1 - \sigma^v)
\Big[\delta_{ij}^{(2)} - k_{ij}
\nonumber \\
&& \quad \; + {i \over 2 k^2} (k_i \Lambda_j^{(v)} - k_j
\Lambda_i^{(v)}) \Big]  + \sigma^v \beta_{ij} \delta({\bf k} \cdot
{\bf B}) \biggr\} \;,
\label{K1}\\
&& \langle b_i \, b_j \rangle^{(0)}({\bf k}) = \langle {\bf b}^2
\rangle^{(0)} \, W^m(k) \, \biggl\{ (1 - \sigma^m)
\Big[\delta_{ij}^{(2)} - k_{ij}
\nonumber \\
&& \quad \; + {i \over 2 k^2} (k_i \Lambda_j^{(m)} - k_j
\Lambda_i^{(m)}) \Big]  + \sigma^m \beta_{ij} \delta({\bf k} \cdot
{\bf B}) \biggr\} \;,
\nonumber \\
\label{K2}
\end{eqnarray}
where $\delta_{ij}^{(2)} = \delta_{ij} - e_i e_j $, $\, \delta_{ij}$
is the Kronecker tensor, $e_i$ is the unit vector which is
perpendicular to the plane of the two-dimensional MHD turbulence,
$k_{ij} = k_i k_j / k^2$, $\, \sigma^v$ and $\sigma^m$ are the
degrees of anisotropy of the velocity and magnetic fluctuations of
the background turbulence, and $\sigma^v > \sigma^m$, $\, \beta_{ij}
=  B_{i}  B_{j} /  B^2 ,$ $\, W^v(k) = W^m(k) = E(k) / 2 \pi k $.
The energy spectrum of the velocity and magnetic fluctuations is
$E(k) = k_{0}^{-1} \, (q-1) \, (k / k_{0})^{-q}$, the turbulent
correlation time is $ \, \tau(k) = 2 \tau_{0} (k / k_{0})^{1-q} ,$
where $1 < q < 3$ is the exponent of the energy spectrum, $ k_{0} =
1 / l_{0} ,$ and $ l_{0} $ is the maximum scale of turbulent
motions, $ \tau_{0} = l_{0} / u_{0}$, $\, u_{0} $ is the
characteristic turbulent velocity in the scale $l_{0}$. The
inhomogeneities of the velocity and magnetic fluctuations of the
background turbulence are characterized by $\, \Lambda_i^{(v)} =
\nabla_i \langle {\bf u}^2 \rangle^{(0)} / \langle {\bf u}^2
\rangle^{(0)}$ and $\Lambda_i^{(b)} = \nabla_i \langle {\bf b}^2
\rangle^{(0)} / \langle {\bf b}^2 \rangle^{(0)}$. Note that $
\langle u_i \, b_j \rangle^{(0)}({\bf k}) = 0$. In Eqs.~(\ref{K1})
and~(\ref{K2}) we neglected small quadratic terms in the parameters
$\Lambda_i^{(v)}$ and $\Lambda_i^{(b)}$.

After the integration in $ {\bf k} $-space we obtain ${\cal E}_{i} =
a_{ij}  B_{j} + b_{ijk}  B_{j,k} ,$ where $ B_{j,k} = \nabla_k
B_{j}$ and
\begin{eqnarray}
a_{ij} &=& 2 \eta_{_{T}} [(1 - \sigma^v) \Lambda_n^{(v)} - \epsilon
(1 - \sigma^m) \Lambda_n^{(m)}] \, \varepsilon_{ipn} K_{pj}^{(1)}
\;,
\nonumber \\
\label{C1}\\
b_{ijk} &=& 2 \eta_{_{T}} \biggl[[1 - \sigma^v + \epsilon (1 -
\sigma^m)] \, (\varepsilon_{ijk} K_{pp}^{(1)} - \varepsilon_{ijp}
K_{pk}^{(1)})
\nonumber \\
& & - 2 \epsilon (1 - \sigma^m) \varepsilon_{ink} K_{nj}^{(1)} + 2
\, \varepsilon_{ink} \tilde K_{nj}^{(1)} [1 - \sigma^v
\nonumber \\
& & - \epsilon (1 - \sigma^m)]  \biggr] + 2 \eta_{_{T}} \beta_{pk}
\, [(\sigma^v + \epsilon \sigma^m) \varepsilon_{ijp}
\nonumber \\
& & - 2 \epsilon \sigma^m \, \varepsilon_{inp} ({\bf e} {\bf \times}
\bec{\hat \beta})_n ({\bf e} {\bf \times} \bec{\hat \beta})_j] \; .
\label{C2}
\end{eqnarray}
Here $\bec{\hat \beta} = {\bf B} / B$,
\begin{eqnarray*}
K_{ij}^{(1)} &=& {1 \over \pi} \int_{{\rm Rm}^{-c}}^{1} x \,
 K_{ij}[y(x)] \,d x \;,
\\
\tilde K_{ij}^{(1)} &=& {1 \over \pi} \int_{{\rm Rm}^{-c}}^{1} \; x
\, y(x) {d  K_{ij}(y) \over dy} \,d x \;,
\\
 K_{ij}(y) &=& \int_0^{2\pi} {k_{ij} \over 1 + y \cos^{2}
\varphi} \,d\varphi =  D_{1}(y) \delta_{ij}^{(2)} +  D_{2}(y)
\beta_{ij},
\\
 D_1(y) &=& {2 \pi \over y} (\sqrt{y + 1} - 1) ,
\; D_2(y) = {2 \pi \over y} \biggl[2 - {y + 2 \over \sqrt{y +
1}}\biggr] ,
\\
y(x) &=& {2 \beta^2 \over x^\gamma} \;, \quad \gamma =  {2 \, (2 -
q) \over q - 1} \;, \quad c = {q-1 \over 3-q} \;,
\end{eqnarray*}
$\beta = 4 \, ( B /  B_{\rm eq})$ and $ B_{\rm eq} = \sqrt{\langle
{\bf u}^2 \rangle^{(0)}}$ is the equipartition field. For $q=5/3$
the parameters $\gamma = 1$ and $c=1/2$, and for $q=3/2$ the
parameters $\gamma = 2$ and $c=1/3$.

To determine the nonlinear turbulent magnetic diffusion
${\eta}_{ij}({\bf B})$ we use an identity: $\eta_{ij} =
(\varepsilon_{ikp} b_{jkp} + \varepsilon_{jkp} b_{ikp}) / 4$. The
nonlinear turbulent magnetic diffusion coefficient along the mean
magnetic field, $\eta_{_{B}}$, and the cross-field turbulent
magnetic diffusion coefficient, $\eta_{_{\perp}}$, are given by:
\begin{eqnarray}
\eta_{_{B}} &=& 2 \eta_{_{T}} \biggl[\sigma^v - \epsilon \sigma^m +
[1 - \sigma^v - \epsilon (1 - \sigma^m)] \Psi_1(\beta) \biggr] \;,
\nonumber \\
\label{B36}\\
\eta_{_{\perp}} &=& 2 \eta_{_{T}} [1 - \sigma^v - \epsilon (1 -
\sigma^m)] \Psi(\beta) \;, \label{B37}
\end{eqnarray}
where $\eta_{_{T}} = l_0 \, u_{0} / 2$, the functions $\Psi(\beta)$,
$\, \Psi_1(\beta)$ and their asymptotic formulas are given in
Appendix B, $\beta = 4 \, ( B / B_{\rm eq})$, $\, B_{\rm eq} =
\sqrt{\langle {\bf u}^2 \rangle^{(0)}}$ is the equipartition field
and the parameter $\epsilon = \langle {\bf b}^2 \rangle^{(0)} /
\langle {\bf u}^2 \rangle^{(0)}$. To derive Eqs.~(\ref{B36}) and
(\ref{B37}) we used the following identities:
\begin{eqnarray*}
&& e_i \varepsilon_{ijp} \beta_{pk} \nabla_k  B_{j} = \Delta_x A \;,
\quad e_i \varepsilon_{ipk} \beta_{pj} \nabla_k  B_{j} = \Delta_y A
\;,
\\
&& e_i \varepsilon_{inp} \beta_{pk} \nabla_k  B_{j} ({\bf e} {\bf
\times} \bec{\hat \beta})_n ({\bf e} {\bf \times} \bec{\hat
\beta})_j = \Delta_x A \;,
\end{eqnarray*}
where ${\bf B} = {\bf e}_x  B$. The nonlinear turbulent magnetic
diffusion coefficients $\eta_{_{B}}$ and $\eta_{_{\perp}}$ for
$\sigma^v=\sigma^m=0$ are given in Sec.~III [see Eqs.~(\ref{BN36})
and (\ref{BN37})].

Now we determine the parameter $\epsilon = \langle {\bf b}^2
\rangle^{(0)} / \langle {\bf u}^2 \rangle^{(0)}$ using budget
equation for the evolution of the mean-square magnetic potential
$\langle a^2 \rangle$ [see Eq.~(\ref{DB32})]. To this end we
determine the energy of magnetic fluctuations $\langle {\bf b}^2
\rangle$ which is obtained from Eq.~(\ref{B24}) by the integration
in $ {\bf k} $-space. The result is given by
\begin{eqnarray}
\langle {\bf b}^2 \rangle &=& {\langle {\bf u}^2 \rangle^{(0)} \over
2} \Big[(1 - \sigma^v) [1 - \phi(\beta)] + \epsilon [1 + \sigma^m
\nonumber \\
&& + \phi(\beta) (1 - \sigma^m)] \Big] \;, \label{DDB33}
\end{eqnarray}
where the function $\phi(\beta)$ and its asymptotic formulas are
given in Appendix B. Combining Eq.~(\ref{DDB33}) with the
steady-state solution of Eq.~(\ref{DB32}) we determine the parameter
$S(\epsilon) \equiv 1 - \sigma^v - \epsilon (1 - \sigma^m)$:
\begin{eqnarray}
S(\epsilon) &=& 2 \biggl[{1 - \sigma^v \over 1 - \sigma^m} + {\rm
Rm} \, {{\rm div} \, {\bf F}^A \over 4 \eta_{_{T}} \, B_{\rm eq}^2}
\biggr] \, \biggl[{1 + \sigma^m \over 1 - \sigma^m} + \phi(\beta)
\nonumber \\
&& + 2 \, {\rm Rm} \, \Psi(\beta) \, { B^2 \over B_{\rm eq}^2}
\biggr]^{-1} \; .
\label{R41}
\end{eqnarray}
When $ B \gg B_{\rm eq} / \sqrt{\rm Rm}$, Eq.~(\ref{R41}) yields the
parameters $\epsilon$:
\begin{eqnarray}
\epsilon &=& {1 \over 1 - \sigma^m} \biggl[1 - \sigma^v  - {{\rm
div} \, {\bf F}^A \over 4 \eta_{_{T}} B^2 \Psi(\beta)} \biggr] \; .
\label{DB41}
\end{eqnarray}
Using Eqs.~(\ref{B36}), (\ref{B37}) and (\ref{DB41}) we obtain the
nonlinear turbulent magnetic diffusion in two directions:
\begin{eqnarray}
\eta_{_{B}} &=& {2 \eta_{_{T}} \over 1 - \sigma^m} \,
\biggl[\sigma^v - \sigma^m + {{\rm div} \, {\bf F}^A \over 4
\eta_{_{T}}  B^2 \Psi(\beta)} [\sigma^m
\nonumber \\
&& + (1 - \sigma^m)\, \Psi_1(\beta)] \biggr] \;,
\label{DB34}\\
\eta_{_{\perp}} &=& {{\rm div} \, {\bf F}^A \over 2  B^2} \; .
\label{DB35}
\end{eqnarray}
Note that there is a small range of the magnitudes of the mean
magnetic field when there can be an anomalous behaviour of the
nonlinear turbulent magnetic diffusion. At $B \to 0.18 \, B_{\rm
eq}$ the function $\Psi(\beta)$ changes sign (see Fig.~1). On the
other hand, the function $S(\epsilon)$ changes sign for slightly
larger value of the magnetic field $B > 0.18 \, B_{\rm eq}$ [see
Eq.~(\ref{R41})]. Therefore, this implies that at $B > 0.18 \,
B_{\rm eq}$ the nonlinear turbulent magnetic diffusion can be
anomalously large. The width of the range of the anomalous behaviour
of the nonlinear turbulent magnetic diffusion is very small, $\delta
B \sim 1 / {\rm Rm}$. In this range the steady-state solution of
Eq.~(\ref{DB32}) for $B > 0.18 \, B_{\rm eq}$ does not exist.

To determine the nonlinear effective drift velocity ${\bf V}^{\rm
eff}({\bf B})$ of the mean magnetic field we use an identity:
$V^{\rm (eff)}_{k} = \varepsilon_{kji} a_{ij} / 2$, which yields
\begin{eqnarray}
{\bf V}^{\rm eff} = - 2 \eta_{_{T}} [(1 - \sigma^v) {\bf
\Lambda}^{(v)} - \epsilon (1 - \sigma^m) {\bf \Lambda}^{(m)}]
\Psi_1(\beta) \;,
\nonumber \\
\label{DB40}
\end{eqnarray}
where the function $\Psi_1(\beta)$ and its asymptotic formulas are
given in Appendix B. When $B \gg  B_{\rm eq} / \sqrt{\rm Rm}$,
Eqs.~(\ref{DB41}) and~(\ref{DB40}) yield
\begin{eqnarray}
{\bf V}^{\rm eff} &=& - 2 \eta_{_{T}} \biggl[(1 - \sigma^v) \, ({\bf
\Lambda}^{(v)} - {\bf \Lambda}^{(m)})
\nonumber \\
&& + {{\rm div} \, {\bf F}^A \over 4 \eta_{_{T}}  B^2 \Psi(\beta)}
\, {\bf \Lambda}^{(m)} \biggr] \Psi_1(\beta) \; .
\label{R43}
\end{eqnarray}
The nonlinear effective drift velocity ${\bf V}^{\rm eff}({\bf B})$
of the mean magnetic field for $\sigma^v=\sigma^m=0$ is given in
Sec.~III [see Eq.~(\ref{RN43})].

\section{Functions $\Psi(\beta)$, $\, \Psi_1(\beta)$ and $\phi(\beta)$}

In this section we present the functions $\Psi(\beta)$, $\,
\Psi_1(\beta)$ and $\phi(\beta)$ used in Sec. III:
\begin{eqnarray*}
\Psi(\beta) &=& {1 \over \pi} \int_{{\rm Rm}^{-c}}^{1} \; x \biggl[1
+ 2 y(x) {d  \over dy} \biggr] \,  [D_1(y) + D_2(y)] \,d x ,
\\
\Psi_1(\beta) &=& {1 \over \pi} \int_{{\rm Rm}^{-c}}^{1} \; x
\biggl[1 + 2 y(x) {d  \over dy} \biggr]  D_1(y) \,d x \;,
\\
\phi(\beta) &=& {1 \over 2 \pi} \int_{{\rm Rm}^{-c}}^{1} \; [2
D_1(y) + D_2(y)] \,d x \; .
\end{eqnarray*}
These functions for $q=5/3$ are given by
\begin{eqnarray}
\Psi(\beta) &=& {\beta^4 \over 6} \, [M(\beta) - M(\beta \, {\rm
Rm}^{1/4})] + {25 \over 2} L(\beta,{\rm Rm})  \;,
\nonumber \\
\label{Y1}\\
\Psi_1(\beta) &=& {1 \over 6 \beta^2} \{2 - [2 - 5 \beta^2 (1 - 3
\beta^2)] \, \sqrt{2 \beta^2 + 1} \}
\nonumber \\
&& + {5 \over 2} L(\beta,{\rm Rm}) \;,
\label{Y2}\\
\phi(\beta) &=& {1 \over 2} \, \biggl[\sqrt{2 \beta^2 + 1} - {2
\over \beta^2} \, L(\beta,{\rm Rm}) \biggr]  \;,
\label{Y5}\\
\nonumber \\
L(\beta,{\rm Rm}) &=& \beta^4 \biggl[ \ln {\sqrt{2 \beta^2 \sqrt{\rm
Rm} + 1} - 1 \over \sqrt{2 \beta^2 \sqrt{\rm Rm} + 1} + 1}
\nonumber \\
&& - \ln {\sqrt{2 \beta^2 + 1} - 1 \over \sqrt{2 \beta^2 + 1} + 1}
 \biggr] \;,
\label{Y4}\\
M(y) &=& {1 \over y^4 \sqrt{2 y^2 + 1}} \, \biggl[5[1 - 5 y^2 (1 +
6 y^2)]
\nonumber \\
&& + {2 \over y^2} (1 - \sqrt{2 y^2 + 1}) \biggr] \; .
\label{Y44}
\end{eqnarray}
Asymptotic formulas for the functions $\Psi(\beta)$, $\,
\Psi_1(\beta)$ and $\phi(\beta)$ are as follows. For $ \beta \ll
{\rm Rm}^{- 1/4} $ these functions are given by
\begin{eqnarray*}
\Psi(\beta) &=& {1 \over 2} \, \biggl[1 - 9 \beta^2 + {25 \over 2}
\beta^4 \ln {\rm Rm} \biggr] \;,
\\
\Psi_1(\beta) &=& {1 \over 2} \, \biggl[1 - 3 \beta^2 + {5 \over
2} \beta^4 \ln {\rm Rm} \biggr] \;,
\\
\phi(\beta) &=& 1 - {1 \over 2} \beta^2 \ln {\rm Rm} \;,
\end{eqnarray*}
for $ {\rm Rm}^{- 1/4} \ll \beta \ll 1 $ they are given by
\begin{eqnarray*}
\Psi(\beta) &=& {1 \over 2} \, \biggl[1 - 9 \beta^2 + 50 \beta^4
|\ln \beta| \biggr] \;,
\\
\Psi_1(\beta) &=& {1 \over 2} \, \biggl[1 - 3 \beta^2 + 10 \beta^4
|\ln \beta| \biggr] \;,
\\
\phi(\beta) &=& 1 - 2 \beta^2 \, |\ln \beta| \;,
\end{eqnarray*}
and for $ \beta \gg 1 $ these functions are given by
\begin{eqnarray*}
\Psi(\beta) &=& - {1 \over 3 \beta^2} \, \biggl[1 - {1.7 \over
\beta} \biggr] \;,
\\
\Psi_1(\beta) &=& {1 \over 3 \beta^2} \;, \quad \phi(\beta) = -
{0.24 \over \beta}  \; .
\end{eqnarray*}

\end{document}